\documentclass[journal]{IEEEtran}
\ifCLASSINFOpdf
\else
\fi
\usepackage[cmex10]{amsmath}
\usepackage{amsmath }
\usepackage{amssymb}
\newtheorem{theorem}{Theorem}
\newtheorem{lemma}{Lemma}
\newtheorem{definition}{Definition}
\newtheorem{construction}{Construction}
\newtheorem{example}{Example}
\usepackage{cite}

\begin{document}
\title{Full-Rank Perfect Codes  over Finite Fields}
%
%
%

\author{Alexander~M.~Romanov~\IEEEmembership{}
\thanks{The work  was supported by Russian Foundation for Basic Research under grant 11-01-00997.}
\thanks{ The author is with the Sobolev Institute of Mathematics,  Siberian Branch of the Russian Academy of Sciences, 4 Academician Koptyug avenue, 630090 Novosibirsk, Russia (e-mail: rom@math.nsc.ru). }}

%
%

\markboth{IEEE TRANSACTIONS ON INFORMATION THEORY}%
{Shell \MakeLowercase{\textit{et al.}}: Bare Demo of IEEEtran.cls for Journals}
%



\maketitle
\begin{abstract}

In this paper, we propose a construction  of full-rank {\boldmath $q$}-ary  1-perfect  codes over finite fields.
This construction is a generalization of the Etzion and Vardy construction   of full-rank binary   1-perfect  codes (1994). Properties of {\boldmath $i$}-components of {\boldmath $q$}-ary  Hamming codes are investigated and the  construction of  full-rank {\boldmath $q$}-ary  1-perfect  codes is based on these properties.
The    switching construction of 1-perfect codes are generalized for the {\boldmath $q$}-ary  case.
We  give  a generalization of the concept of {\boldmath $i$}-component of 1-perfect codes and introduce the concept of {\boldmath $(i,\sigma)$}-components of {\boldmath $q$}-ary 1-perfect codes.
We  also present a generalization of the Lindstr\"{o}m and Sch\"{o}nheim  construction of {\boldmath $q$}-ary 1-perfect  codes and provide a  lower bound on the number  of pairwise distinct {\boldmath $q$}-ary 1-perfect codes of length {\boldmath $n$}.

\end{abstract}


%
\IEEEpeerreviewmaketitle

\section {Introduction}

\IEEEPARstart{L}{et} $\mathbb{F}_{q}^n$ be a vector space of dimension $n$ over the finite field $\mathbb{F}_{q}$ of order $q$.
An arbitrary subset  $\cal {C}$ of  $\mathbb{F}_{q}^n$ is called a $q$-ary  {\it  code} of length $n$.
The vectors in $\cal{C}$ are called {\it codewords}. A code  $\cal {C}$ is called {\it linear} if it is a linear space over $\mathbb{F}_{q}$.
The orthogonal complement ${\cal C}^{\perp}$ is the {\it dual} code of ${\cal C}$.
A {\it generator matrix} of a linear code $\cal{C}$ is matrix  whose rows form a basis for $\mathcal{C}$.
A {\it parity-check matrix} of a linear code $\cal{C}$ is matrix  whose rows form a basis for $\cal{C}^{\bot}$.
The { \it (Hamming) distance} between two vectors ${\bf x}$, ${\bf y} \in \mathbb{F}_{q}^n$ is the number of coordinates in which they differ, and it is denoted by $d({\bf x}, {\bf y})$.
The {\it (Hamming) weight} $wt({\bf x})$ of a vector ${\bf x} \in \mathbb{F}_{q}^n$ is the number of nonzero coordinates in ${\bf x}$.
The  {\it (minimum)} {\it distance}  $d({\cal C})$ of a code ${\cal C}$ is the smallest distance between distinct codewords.
The linear code ${\cal C} \subseteq  \mathbb{F}_{q}^n$  has parameters $[n, k, d]_q$ if ${\cal C}$ has length $n$, dimension $k$, and minimum distance  $d$.
The nonlinear code ${\cal C} \subseteq  \mathbb{F}_{q}^n$  has parameters $(n, |{\cal C}|, d)_q$ if ${\cal C}$ has length $n$, number of codewords $|{\cal C}|$, and minimum distance  $d$.

We define a map $ p : \mathbb{F}_{q}^n \rightarrow \mathbb{F}_{q} $.  If ${\bf x}= (x_1, x_2, \ldots, x_n) \in \mathbb{F}_{q}^n$ then $p({\bf x}) = \sum_{i = 1}^{n}x_i $.
If ${\cal C}$ is an $(n, |{\cal C}|, d)_q$ code over $F_q$  then  the {\it extended code} ${\widehat {\cal C}}$ to be the code

$${\widehat {\cal C}}=\{ ({\bf x}  \  |  \  p({\bf x})): {\bf x} \in {\cal C}\}. $$
The extended code ${\widehat{\cal C}}$ is an $(  n + 1, |{\cal C}|, {\hat d} )_q$ code, where ${\hat d}$ equals $d$ or $d + 1$.

If  $\cal {C}$  is a $q$-ary $(n,|{\cal C}|, d = 2t + 1)_q$ code, then we have

\begin{equation}
|{\cal C}|\sum_{k = 0}^t \binom{n}{k}(q-1)^k \leq q^n.
\end{equation}

If the equality  occurs, then ${\cal C}$ is called a {\it perfect code}. The inequality (1) is known as the  Sphere-Packing Bound.

A $q$-ary  perfect code ${\cal C}$ with parameters $(n,|{\cal C}|, 3)_q$ is called a $q$-ary {\it 1-perfect}.
It is known that the $q$-ary 1-perfect  codes of length $n$   exist only if $n = (q ^{m} -1) / (q-1)$, where $m$ is a natural number not less than two.

A mapping $\mathbb{F}_{q} \rightarrow  \mathbb{F}_{q}$ is called an {\it isometry}, if it preserves distances.
It is known that two codes ${\cal C}_1, {\cal C}_2 \subseteq \mathbb{F}_{q}^n $ are  {\it isometric} if and only if there are $n$ permutations
$\sigma_1,  \sigma_2, \ldots, \sigma_n$ of the elements in $\mathbb{F}_{q}$ and permutation $\pi$ of the coordinates such
that ${\cal C}_2 =\{\pi(\sigma_1(c_1), \sigma_2(c_2), \ldots, \sigma_n(c_n) ) :  (c_1, c_2, \ldots, c_n) \in {\cal C}_1  \}$.

Two codes ${\cal C}_1, {\cal C}_2 \subseteq \mathbb{F}_{q}^n $ to be {\it equivalent} if there exists a vector ${\bf v} \in \mathbb{F}_{q}^n $ and $n \times n$ monomial matrix $M$ over $\mathbb{F}_{q}$ such that ${\cal C}_2 =\{({\bf v} + {\bf c}M) :  {\bf c} \in  {\cal C}_1  \}$.

In binary case, two codes are equivalent if and only if they are isometric. In nonbinary case, if  $q \geq 5$ then a linear  $q$-ary  1-perfect code can be isometric to a nonlinear  $q$-ary  1-perfect code.

A linear $q$-ary  1-perfect code of length $n$ is unique up to equivalence and is called a $q$-ary
{\it Hamming code}.
We will denote the  $q$-ary  Hamming code of length $n = (q^m - 1) / (q - 1)$  by ${\cal H}_{q,m}$.

Besides the Hamming codes, there are also linear perfect codes that are called the {\it Golay codes}. The  Golay codes are unique up to equivalence.  The Nonlinear perfect codes with parameters of the Golay codes do not exist. The Golay codes are listed in Table I.

The {\it trivial} perfect codes include codes that have only one codeword,  codes that are the whole of $\mathbb{F}_{q}^n $ and binary repetition codes of odd length that consisting of only two codewords, namely, the all one vector and the all zero vector.

It is known \cite{lint, ti, zin}  that any non-trivial perfect code  over the finite field has the parameters of a Hamming code or a Golay code.

The {\it rank} of a code $\cal C$ is the maximum number of linearly independent codewords of $\cal C$.  A code of length $n$ that has  rank $n$ is said to have full rank; otherwise, the code is non-full rank.

It is known \cite{vas1, lin, sch} that there exist at least $q^{q^{cn}}$ (pairwise) nonequivalent $q$-ary  1-perfect codes of length $n$ where $c = \frac{1}{q} - \epsilon $.
If $q = 2$ and $n = 15$, then there exist exactly $ 5 \, 983$ nonequivalent binary  1-perfect codes of length $15$, see \cite{ost1}.

In this paper, we propose a construction  of full-rank $q$-ary  1-perfect  codes  over finite fields. This construction is a generalization of the  Etzion and Vardy  construction   of full-rank binary   1-perfect  codes \cite{etz}. Properties of $i$-components of $q$-ary  Hamming codes are investigated and the  construction of  full-rank $q$-ary  1-perfect  codes is based on these properties.
The   switching construction (Vasil'ev \cite{vas1}, Etzion and Vardy \cite{etz}) of 1-perfect codes  are generalized for the $q$-ary  1-perfect  codes.
We  give  a generalization of the concept of $i$-component of 1-perfect codes and  introduce the concept of $(i,\sigma)$-components of 1-perfect $q$-ary codes.
We  also present a generalization of the Lindstr\"{o}m \cite{lin} and Sch\"{o}nheim  \cite{sch}  construction of $q$-ary 1-perfect  codes and provide a  lower bound on the number  of (pairwise distinct) $q$-ary 1-perfect
codes of length $n$.

In Section II we present the  construction of Lindstr\"{o}m \cite{lin} and Sch\"{o}nheim  \cite{sch}.
In Section III we  give  a generalization of the concept of $i$-component and   introduce the concept of $(i,\sigma)$-components.  In Section III we also  present the generalization of the switching construction.
In Section IV we investigate properties of $i$-components of $q$-ary  Hamming codes, we also  present a   generalization of the Lindstr\"{o}m \cite{lin} and Sch\"{o}nheim  \cite{sch}  construction and  provide a lower bound on the number  of  $q$-ary 1-perfect codes of length $n$.
Section V  presents  the  construction of full-rank $q$-ary  1-perfect  codes  over finite fields. We close in
Section VI with some remarks.

\begin{table}[!t]
\renewcommand{\arraystretch}{1.3}
\caption{Golay Codes}
\label{table}
\centering
\begin{tabular}{c|c|c|c|l}

       $q$ & $n$ & $|{\cal C}|$ & $d({\cal C})$ & { name of code ${\cal C}$ } \\  \hline

       $2$ & $23$ & $4096$ & $7$ & { binary Golay code}\\

       $2$ & $24$ & $4096$ & $8$ & { extended binary Golay code} \\

       $3$ & $11$ & $729$ & $5$ & { ternary Golay code}   \\

       $3$ & $12$ & $729$ & $6$ &  { extended ternary Golay code} \\

\end{tabular}
\end{table}

\section {The $(\bf u | \bf u + \bf v)$ construction and generalizations}

In this section, we present  the  construction of Lindstr\"{o}m \cite{lin} and Sch\"{o}nheim  \cite{sch} of  $q$-ary 1-perfect codes.
We will start from the well-known $(\bf u | \bf u + \bf v)$ construction.

Two codes of the same length can be combined to form a third code of twice the length.
Let $ {\cal C}_i$ be an $[n, k_i , d_i ]_q$ code for $i \in \{1, 2\}$, both over the same finite field $\mathbb{F}_{q}$.
The $({\bf u} | {\bf u} + {\bf v})$ construction produces the $[2n, k_1 + k_2, min \{2d_1, d_2\}]_q$ code

$${\cal C} = \{({\bf u} | {\bf u} + {\bf v}) : {\bf u} \in {\cal C}_1,{\bf v} \in {\cal C}_2\}.$$

Let $\widehat {\cal H}_{2,m}$ be a extended binary   Hamming code of length $  2^m = n+1$. Then
$$\widehat {\cal H}_{2,m +1} = \{({\bf u}| {\bf u} + {\bf v}) : {\bf u} \in \mathbb{F}_{2}^{n + 1}, p({\bf u}) = 0, {\bf v} \in \widehat {\cal H}_{2,m} \}$$
is the extended binary  Hamming code of length $ 2^{m + 1} = 2n+2$.
Let $ {\cal H}_{2,m}$ be a binary Hamming code of length $2^m -1 = n$.
Then
$${\cal H}_{2,m + 1} = \{({\bf u}| {\bf u} + {\bf v}|p({\bf u})) : {\bf u} \in \mathbb{F}_{2}^{n}, {\bf v} \in {\cal H}_{2,m} \}$$
is the  binary Hamming code of length $ 2^{m + 1} - 1 = 2n+1  $.


\begin{construction}[ Vasil'ev \cite{vas1}]
Given a binary 1-perfect code ${\cal C}_1 \subset \mathbb{F}_{2}^n$ and function $\lambda$ defined on ${\cal C}_1$ with values in $\mathbb{F}_{2}$, we construct the code ${\cal C} \subset  \mathbb{F}_{2}^{2n +1} $. Let
$$ {\cal C} = \left\{( {\bf u}  |  {\bf u} + {\bf v}    |   p({\bf u}) + \lambda({\bf v}) ):
 {\bf u} \in  \mathbb{F}_{2}^n, {\bf v} \in {\cal C}_1 \right\}. $$
The code ${\cal C}$ is a binary 1-perfect code of length $2n + 1$. If function $\lambda$ is nonlinear then  the 1-perfect code ${\cal C}$ is nonlinear.
\end{construction}


Next, we describe so called doubling construction of binary 1-perfect codes (or extended binary 1-perfect codes). The  doubling construction is a combinatorial generalization of the $(\bf u | \bf u + \bf v)$ construction for the binary 1-perfect codes.

First we give definition of a map $\varphi$.
Let ${\cal C}_0, {\cal C}_1, \ldots, {\cal C}_{n}$ be a partition of $\mathbb{F}_{2}^n$ into binary  1-perfect codes of length  $n$. We   denote by $Q$ the set $\{0, 1, \ldots, n\}$ and define a map
$ \varphi : \mathbb{F}_{2}^n \rightarrow Q$.
If ${\bf u} \in {\cal C}_k$, then $\varphi({\bf u}) = k$, where $k \in Q$.


\begin{construction}[Doubling Construction]
Let $ {\cal C}_0^1, {\cal C}_1^1, \ldots, {\cal C}_n^1$ and  ${\cal C}_0^2, {\cal C}_1^2, \ldots, {\cal C}_n^2$ be two partitions of $\mathbb{F}_{2}^n$ into binary 1-perfect codes of length  $n$ and
$\pi : Q \rightarrow Q$ be a permutation of the set  $Q$.
Then
\setlength{\arraycolsep}{0.0em}
\begin{eqnarray}
\nonumber {\cal C}  &{} = {}&  \bigl\{( {\bf u}  |  {\bf v}    |   p({\bf u}) ): \\
\nonumber & & {\bf u} \in  \mathbb{F}_{q}^n = {\cal C}_0^1 \cup {\cal C}_1^1 \cup \cdots \cup {\cal C}_n^1 \; \, \mbox{and} \; \,  {\bf v} \! \in {\cal C}_{\pi(\varphi({\bf u}))}^2 \bigr\}.
\end{eqnarray}
\setlength{\arraycolsep}{5pt}
It is obvious, that  ${\cal C} $ is a binary 1-perfect code of length $2n + 1$.
\end{construction}


Heden \cite{hed1} constructed by the doubling construction a nonlinear binary 1-perfect code of length $n = 15$ that is not equivalent to the Vasil'ev codes. Phelps \cite{ph1} and  Solov'eva \cite{sol} constructed  partitions of $\mathbb{F}_{2}^n$ into binary 1-perfect codes of length  $n$, that is not partitions into  translations of a 1-perfect code or partitions into  cosets of a Hamming code.

Lindstr\"{o}m \cite{lin}  and  Sch\"{o}nheim  \cite{sch} have generalized the construction of Vasil'ev  to the $q$-ary 1-perfect codes.


\begin{construction}[Lindstr\"{o}m \cite{lin}, Sch\"{o}nheim  \cite{sch}]
Given a $q$-ary 1-perfect code ${\cal C}_1 \subset \mathbb{F}_{q}^n$ and function $\lambda$ defined on ${\cal C}_1$ with values in $\mathbb{F}_{q}$ we construct the code ${\cal C} \subset  \mathbb{F}_{q}^{qn +1} $. Let $\alpha_1, \alpha_2, \ldots, \alpha_{q - 1}$ be all nonzero elements in  $\mathbb{F}_{q}$. Then
\setlength{\arraycolsep}{0.0em}
\begin{eqnarray}
\nonumber {\cal C} & {} = {} \Biggl\{&\Biggl ({\bf u}_1 | {\bf u}_2 | \cdots | {\bf u}_{q-1} |{\bf v }   +  \sum_{i =1}^{q-1} {\bf u}_i |
\sum_{i =1}^{q-1}\alpha_i  p({\bf u}_i) + \lambda({\bf v}) \Biggr)\!: \\
\nonumber& & {\bf u}_i \in   \mathbb{F}_{q}^n, \alpha_i \in  \mathbb{F}_{q} \setminus \{{ 0}\},  i \in \{1,  \ldots, q-1\},  {\bf v} \in {\cal C}_1 \Biggr\} .
\end{eqnarray}
The code ${\cal C} $ is a $q$-ary 1-perfect code of length $qn + 1$.
\end{construction}


\section {The $(i,\sigma)$-components}

In this section, we  give  a generalization of the concept of $i$-component and   introduce the concept of $(i,\sigma)$-components and we also  present the generalization of  switching construction.

Let $\cal C$ be a code over $\mathbb{F}_{q}$. A {\it distance graph} of the code ${\cal C}$ is a graph whose vertex set is ${\cal C}$ and vertices
${\bf x}, {\bf y} \in  {\cal C}$ are adjacent if and only if $d({\bf x}, {\bf y}) = d$, where $d$ is a fixed  natural number. If $d$ is minimum distance of the code ${\cal C}$, then the distance graph is called {\it minimum distance graph}.
We can puncture $\cal C$ by deleting the same coordinate $i$ in each codeword.
We  denote the punctured code by ${\cal C}^{i}$.

Next, we give the   Vasil'ev  definition  of an $i$-component of  binary 1-perfect code ${\cal C}$  of length $n$, where a coordinate $i \in \{1, 2, \ldots, n\}$.


\begin{definition}[Vasil'ev \cite{vas2}]
An $i$-component of a code ${\cal C}$ is a subcode (or subset) of the ${\cal C}$ and codewords of $i$-component of   ${\cal C}$ correspond to  vertices of connected component of  minimum distance graph of the punctured code ${\cal C}^i$. The set of all $i$-components of   ${\cal C}$ correspond to the set of all connected components of  minimum distance graph of the punctured code ${\cal C}^i$.
\end{definition}

\begin{example} Consider the binary Hamming code
$${\cal H}_{2,3} = \{( {\bf u}|{\bf u} + {\bf v}| p({\bf u})) : {\bf u} \in \mathbb{F}_{2}^3,  {\bf v} \in \{(0, 0, 0), (1, 1, 1)\}\}$$ of length $7$. Suppose that coordinate $i = 7$. Then,  the binary Hamming code ${\cal H}_{2,3}$ has two $7$-components:
\setlength{\arraycolsep}{0.0em}
\begin{eqnarray}
 \nonumber{\cal R}_7  &{}={} &
\left\{\begin{array}{ccccccc}
               (0, & 0, & 0,  & 0, & 0, & 0, & 0) \\
               (0, & 0, & 1,  & 0, & 0, & 1, & 1) \\
               (0, & 1, & 0,  & 0, & 1, & 0, & 1) \\
               (0, & 1, & 1,  & 0, & 1, & 1, & 0) \\
               (1, & 0, & 0,  & 1, & 0, & 0, & 1) \\
               (1, & 0, & 1,  & 1, & 0, & 1, & 0) \\
               (1, & 1, & 0,  & 1, & 1, & 0, & 0) \\
               (1, & 1, & 1,  & 1, & 1, & 1, & 1)
\end{array}\right\}, \\
 \nonumber\phantom{and} & & \\
{\cal R}_7 + (0, 0, 0,  1, 1, 1, 0)  &{}={}&
 \nonumber\left\{\begin{array}{ccccccc}
               (0, & 0, & 0,  & 1, & 1, & 1, & 0) \\
               (0, & 0, & 1,  & 1, & 1, & 0, & 1) \\
               (0, & 1, & 0,  & 1, & 0, & 1, & 1) \\
               (0, & 1, & 1,  & 1, & 0, & 0, & 0) \\
               (1, & 0, & 0,  & 0, & 1, & 1, & 1) \\
               (1, & 0, & 1,  & 0, & 1, & 0, & 0) \\
               (1, & 1, & 0,  & 0, & 0, & 1, & 0) \\
               (1, & 1, & 1,  & 0, & 0, & 0, & 1)
\end{array}\right\}.
\end{eqnarray}
\setlength{\arraycolsep}{5pt}
\end{example}
A binary Hamming code of length $7$ has two $i$-components for each coordinate $i \in \{1, 2, \ldots, 7\}$.



It is not difficult to generalize the definition  of Vasil'ev to $q$-ary 1-perfect codes.
Further we  give  a generalization of the  switching construction for arbitrary $q$-ary 1-perfect codes.

Given an $i$-component  ${\cal R}_i \subseteq \mathbb{F}_{q}^n$ and  permutation $\sigma$ of the elements in $\mathbb{F}_{q}$, we define  the code $\sigma({\cal R}_i)$. Let
\setlength{\arraycolsep}{0.0em}
\begin{eqnarray}
\nonumber \sigma({\cal R}_i)&{}={}& \{({x}_1, {x}_2, \ldots, \sigma({x}_i),   \ldots, {x}_n ):  \\
\nonumber & & \phantom{ a a a a a a a a a a }  ({x}_1, {x}_2, \ldots, {x}_i,   \ldots, {x}_n ) \in {\cal R}_i \}.
\end{eqnarray}
\setlength{\arraycolsep}{5pt}


\begin{construction}
Let  $ \{ {\cal R}_{i}(1), {\cal R}_{i}(2), \ldots, {\cal R}_{i}(t)  \}$ be a family of $i$-components of a $q$-ary 1-perfect code ${\cal C}_1$ of length $n$ and  let  $\sigma_{1},  \sigma_{2}, \ldots, \sigma_{t}$ be $t$ permutations of the elements in $\mathbb{F}_{q}$. Then we construct the code ${\cal C}$. Let
\begin{equation}
{\cal C} = \left({\cal C}_1 \setminus  \bigcup_{s = 1}^t {\cal R}_{i}(s) \right) \bigcup \left( \bigcup_{s = 1}^t \sigma_{s}({\cal R}_{i}(s)) \right).
\end{equation}
\end{construction}


\begin{theorem}
The code ${\cal C}$ is a $q$-ary 1-perfect code of length $n$.
\end{theorem}

\begin{IEEEproof}

We need to prove that the number of codewords in the code ${\cal C}$ is correct and  that the minimum distance $d({\cal C})$ of the code ${\cal C}$ is equal to $3$. Obviously that $|{\cal R}_{i}(s)|= |\sigma_{s}({\cal R}_{i}(s))|$. By definition of $i$-components, it follows that  the $i$-components from $ \{ {\cal R}_{i}(1), {\cal R}_{i}(2), \ldots, {\cal R}_{i}(t)  \}$ pairwise disjoint.  Thus we have
$$|{\cal C}| = |{\cal C}_1| - \sum_{s = 1}^t|{\cal R}_{i}(s)| +  \sum_{s = 1}^t|\sigma_{s}({\cal R}_{i}(s))| = |{\cal C}_1|
= q^{n - m}. $$
The permutation $\sigma_{s}$ defines  an isometric transformation on the space $\mathbb{F}_{q}^n$. Hence we have
$d(\sigma_{s}({\cal R}_{i}(s))) = 3$, $s \in \{1, 2, \ldots, t\}$. Consider  punctured codes $({\cal R}_{i}(s))^{i}$ and
$({\cal C}_1  \setminus {\cal R}_{i}(s))^{i}$.
It follows from the definition of the $i$-components of the code  that for any ${\bf x}\in ({\cal R}_{i}(s))^{i}$ and for any ${\bf y}\in ({\cal C}_1  \setminus {\cal R}_{i}(s))^{i}$ there holds the inequality $d({\bf x},{\bf y}) \geq 3$. Hence   for any ${\bf x}\in \sigma_{s}({\cal R}_{i}(s))$ and for any ${\bf y}\in ({\cal C}_1  \setminus {\cal R}_{i}(s))$ we have $d({\bf x},{\bf y}) \geq 3$.
Since the $i$-components from $ \{ {\cal R}_{i}(1), {\cal R}_{i}(2), \ldots, {\cal R}_{i}(t)  \}$ pairwise disjoint, we obtain that the code ${\cal C}$ is a $q$-ary 1-perfect code of length $n$.
\end{IEEEproof}

In binary case, there exists a  unique nontrivial permutation of the elements in $\mathbb{F}_{2}$ and  formula (2) takes the form
$$ {\cal C} = \left({\cal C}_1 \setminus  \bigcup_{s = 1}^t {\cal R}_{i_s} \right) \bigcup \left( \bigcup_{s = 1}^t {\cal R}_{i_s} + {\bf e}_{i_s} \right)$$
where ${\bf e}_{i_s}$ is a  vector in which $i_s$-th component is equal to $1$ and other components are equal to $0$.

We now give  a generalization of the concept of $i$-component of 1-perfect codes and introduce the concept of $(i,\sigma)$-components of $q$-ary 1-perfect  codes.  Particular  cases of the $(i,\sigma)$-components  were considered in \cite{rom1}, \cite{ph5}. We  present  two definition of $(i,\sigma)$-components.
Following \cite{rom1}, we give a recursive definition of the $(i,\sigma)$-components.


\begin{definition}
Given a  $q$-ary 1-perfect  code  ${\cal C} \subset \mathbb{F}_{q}^n$, coordinate $i$, and  permutation $\sigma$ of the elements in $\mathbb{F}_{q}$, a subcode ${\cal R}_{(i,\sigma)} \subseteq{\cal C}$ is an $(i,\sigma)$-component of the code ${\cal C}$
if    ${\bf y} \in {\cal R}_{(i,\sigma)}$ implies that
$$\{{\bf x}: {\bf x} \in {\cal C}, d({\bf x }, {\bf y}(i, \sigma)) = 2 \} \subseteq {\cal R}_{(i,\sigma)}$$
where
$$ {\bf y}(i, \sigma) = (y_1, y_2, \ldots, \sigma(y_i), \ldots, y_n). $$
\end{definition}

Following \cite{ph5}, we formulate the definition of the $(i,\sigma)$-components in terms of   graph theory.
Given a $q$-ary  code  ${\cal C} \subseteq \mathbb{F}_{q}^n$, coordinate $i$ and  permutation $\sigma$ of the elements in $\mathbb{F}_{q}$, we define  the code ${\cal C}(i,\sigma)$. Let
\setlength{\arraycolsep}{0.0em}
\begin{eqnarray}
 \nonumber   {\cal C}(i,\sigma) & {}={} & \{({x}_1, {x}_2, \ldots, \sigma({x}_i), \ldots, {x}_n ): \\
 \nonumber & & \phantom{ a a a a a a a a a a }  ({x}_1, {x}_2, \ldots, {x}_i,   \ldots, {x}_n ) \in {\cal C}\}.
\end{eqnarray}
\setlength{\arraycolsep}{5pt}
\begin{definition}
 Given a $q$-ary 1-perfect code ${\cal C} \subset \mathbb{F}_{q}^n$, coordinate $i$, and
permutation $\sigma$ of the elements in $\mathbb{F}_{q}$, we consider the distance bipartite graph of the code ${\cal C}\cup {\cal C}(i,\sigma)$. Two codewords ${\bf x} \in {\cal C}$ and ${\bf y} \in {\cal C}(i,\sigma)$ are adjacent if and only if $d({\bf x}, {\bf y}) = 2$. The $(i,\sigma)$-components of the code ${\cal C}$ correspond to  the connected components of the distance bipartite graph of the code ${\cal C}\cup {\cal C}(i,\sigma)$.
\end{definition}


Let ${\cal R}_{i,\sigma}$ be an $(i,\sigma)$-component of a $q$-ary 1-perfect  code ${\cal C}_1$ of length $n$ and let
\setlength{\arraycolsep}{0.0em}
\begin{eqnarray}
\nonumber  \sigma({\cal R}_{i,\sigma})&{}={}& \{({x}_1, {x}_2, \ldots, \sigma({x}_i),   \ldots, {x}_n ): \\
\nonumber& & \phantom{ a a a a a a a a a a } ({x}_1, {x}_2, \ldots, {x}_i,   \ldots, {x}_n ) \in {\cal R}_{i,\sigma}\}.
\end{eqnarray}
\setlength{\arraycolsep}{5pt}
Then, it is obvious that
$${\cal C} = \left({\cal C}_1 \setminus  {\cal R}_{i,\sigma} \right) \cup \left( \sigma({\cal R}_{i,\sigma}) \right) $$
is a $q$-ary 1-perfect code of length $n$.

In binary case, an $(i,\sigma)$-component is an $i$-component.
In nonbinary case,  $(i,\sigma)$-components form a partition of an $i$-component.

Let ${\cal C}_1$ be a $q$-ary 1-perfect code, ${\cal R}_{i,\sigma}$ be an $(i,\sigma)$-component of  ${\cal C}_1$ and let
$${\cal C}_2 = \left({\cal C}_1 \setminus  {\cal R}_{i,\sigma} \right)
\cup \left( \sigma({\cal R}_{i,\sigma}) \right).$$
Then  we say that the code ${\cal C}_2$ is obtained from  ${\cal C}_1$ by a switching $(i,\sigma)$-component ${\cal R}_{i,\sigma}$.

Let ${\cal C}_1, {\cal C}_2, \ldots, {\cal C}_t$ be the $q$-ary 1-perfect codes of length $n$ and let code $C_{s + 1}$  be
obtained from $C_{s}$ by a switching $(i_s,\sigma_s)$-component, where $s \in \{1, 2, \ldots, t - 1 \}$, $i_s \in \{1, 2, \ldots, n \}$, $\sigma_s$ is a permutation  of the elements in $\mathbb{F}_{q}$. Then we say that ${\cal C}_t $ is obtained from  ${\cal C}_1 $ by a sequence of switchings.

The {\it switching class} of a $q$-ary 1-perfect code ${\cal C }$ is a collection of all  nonequivalent $q$-ary  1-perfect codes that can be obtained from ${\cal C }$ by a sequence of switchings.

A full-rank $q$-ary  1-perfect  code is called Type I, if its switching class contains  non-full-rank codes; otherwise, the code is called Type II.

The problem of existence of Type II full-rank  $q$-ary  1-perfect codes   is  open.
Originally, this problem was posed for binary codes in \cite{rom5}.

It is known \cite{ost2}  that there are $9$ switching classes for the binary 1-perfect codes
of length $15$, and their sizes are $5 819$, $153$, $3$, $2$, $2$, $1$, $1$, $1$, and $1$. The switching class
of the Hamming code have $5 819$ nonequivalent codes and in fact contains all codes with full rank except two. The two full-rank codes that are not in the switching class
of the Hamming code have one more code in their switching
class, a code with rank $14$.
Consequently, the Type II full-rank   binary 1-perfect codes of length $15$ do not exist.

Phelps and LeVan \cite{ph2} constructed, by the doubling construction, a binary 1-perfect code of length $15$  whose switching class consists of just two nonequivalent codes.
Etzion and Vardy \cite{etz} showed  that full-rank binary 1-perfect codes  can  not be constructed by doubling construction.
Heden and Krotov \cite{hed2} showed  that  non-full-rank $q$-ary 1-perfect codes have certain structural properties.

\section {Properties of $i$-components}

In this section, we investigate properties of  $i$-components of $q$-ary  Hamming codes, we also  present a   generalization of the Lindstr\"{o}m \cite{lin} and Sch\"{o}nheim  \cite{sch}  construction and  provide  a lower bound on the number  of  $q$-ary 1-perfect codes of length $n$. At first  we prove theorems describing some properties of  $i$-components of $q$-ary Hamming  code ${\cal H}_{q,m}$.

The parity-check matrix $H=\left[{\bf h}_1, {\bf h}_2,  \ldots,  {\bf h}_n\right]$ of the code ${\cal H}_{q,m}$ of length $n = (q^{m} - 1)/(q-1)$ consists of $n$ pairwise linearly independent column  vectors ${\bf h}_i$, $i \in \{1, \dots, n \}$.
The transposed column vector ${\bf h}_i^T$ belongs to $\mathbb {F}_{q}^{m}$, $i \in \{1, \dots, n \}$. We assume that the columns of the parity-check matrix $H$ are arranged in some fixed order. The set  $\mathbb{F}_{q}^{m} \setminus \{ \bf 0 \}$ generates a projective geometry $PG_{m-1} (q)$ of dimension $(m-1)$ over the finite  field  ${\mathbb{F}}_{q}$. In this geometry, points correspond to the columns of the parity-check matrix  $H$ and the three points $i, j, k$ lie on the same line if the corresponding columns  $\bf h_i, \bf h_j, \bf h_k$ are linearly dependent.
We denote by $l_{xy}$ the line passing through the points $x$ and $y$, and we denote by $P_{xyz}$ the plane spanned by three non-collinear points $x, y, z$. Let ${\bf x} = (x_1, x_2, \dots, x_n) \in \mathbb{F}_{q}^{n}$, then  the {\it support} of the vector ${\bf x}$ is the set $supp({\bf x}) = \{ i : x_i \neq 0 \}$.
A triple belongs to a line $l$ if the support of this triple belongs to the line $l$. The triples intersect at a point $i$ if their supports intersect at the point $i$.

A vector of weight $3$ of the $q$-ary  Hamming  code ${\cal H}_{q,m}$ is called {\it triple}. Following \cite{ph3, ph4}, we denote by  ${\cal R}_i$  a  subspace spanned by the set of all triples of the code  ${\cal H}_{q,m}$  having  $1$ in the  $i$-th coordinate.

Consider a vector ${\bf x} \in  \mathbb{F}_{q}^{n}$
such that its $supp({\bf x})$ is $m - 2$ dimensional hyperplane. Denote
by  $\mathbb{F}_{q}^{n}({\bf x})$ the set of all vectors ${\bf u} \in  \mathbb{F}_{q}^{n}$ such that $supp({\bf u}) \subseteq supp({\bf x})$.

Denote by  ${\cal H}_l$ the subcode  of  ${\cal H}_{q,m}$  defined by a line  $l$. We remind that, by definition,
$${\cal H}_l =\{{\bf u}: {\bf u} \in {\cal H}_{q,m} \; \mbox{and} \; supp({\bf u}) \subseteq l \}.$$

\begin{lemma}[Romanov \cite{rom7}, Lemma 1]
Let $i \notin supp ({\bf x})$ and ${\bf u} \in  \mathbb{F}_{q}^{n} ({\bf x})$. Then, the intersection
$$({\cal R}_i + {\bf u}) \cap \mathbb{F}_{q}^{n}({\bf x})$$
contains only one vector.
\end{lemma}

\begin{IEEEproof}
Consider a pencil of lines $l_1, l_2, \dots, l_{(n-1)/q}$ which pass through  the point $i$. It is known \cite{rom4} that
\begin{equation}
{\cal R}_i = {\cal H}_{l_1} +  {\cal H}_{l_2} + \cdots + {\cal H}_{l_{(n-1)/q}}. \label{eq1}
\end{equation}
Let $l_s$ be an arbitrary line through the point $i$, where $s \in \{1, 2, \ldots, (n -1)/q\}$. Since $i \notin supp({\bf x})$, it follows that any line passing through the point $i$ intersects with the hyperplane $supp({\bf x})$ only at one point. We can consider projective geometry $PG_{m-1} (q)$ as the set of
all subspaces of the vector space $\mathbb{F}_{q}^{n}$. A point in projective geometry is a subspace of dimension one.  Hence the intersection of $ {\cal H}_{l_s} \cap \mathbb{F}_{q}^{n}({\bf x})$ can contain only  vectors of weight 0 or 1. It is obvious that ${\bf 0} \in  {\cal H}_{l_s} \cap \mathbb{F}_{q}^{n}({\bf x})$. Since the minimum weight of the nonzero vectors in ${\cal H}_{l_s}$ is equal to $3$, it follows that ${\cal H}_{l_s} \cap \mathbb{F}_{q}^{n}({\bf x}) = \{{ \bf 0} \}$. Since the line $ l_s $ was chosen arbitrarily  and  $ {\cal R}_i^n$ is a subspace, we get that
${\cal R}_i^n \cap  \mathbb{F}_{q}^{n}({\bf x}) = \{ {\bf 0} \}$.
\end{IEEEproof}

Each line through the point $i$ contains $q -1$ linearly independent triples. Therefore from (3) we obtain  that
the dimension of ${\cal R}_i$  is equal to $(q -1)({(n-1)/q}) =  q^{m -1} - 1$ see \cite{ph4}.

\begin{theorem}
Given a coordinate $i$ and ${\bf u} \in   {\cal H}_{q,m}$, a code ${\cal R}_i + {\bf u}$ is an  $i$-component of the $q$-ary  Hamming code ${\cal H}_{q,m}$ of length $n = (q^m - 1) / (q - 1), \ m \geq 2$.
\end{theorem}


\begin{IEEEproof}
Consider  punctured code ${\cal R}_{i}^{i}$. Obviously that code ${\cal R}_{i}^{i}$ is  linear and spanned by codewords of weight $2$. 	Thus the code  ${\cal R}_{i}^{i}$ is  Hamiltonian and minimum distance  graph of the code ${\cal R}_{i}^{i}$ is connected, see \cite{rom7}. Without loss of generality we can assume that $i \notin supp ({\bf x})$, where  ${\bf x} \in  \mathbb{F}_{q}^{n}$ and $supp({\bf x})$ is a $m - 2$ dimensional hyperplane.
Hence taking into account Lemma 1 we obtain  that
$${\cal H}_{q,m} = \bigcup_{{\bf u}\in {\cal H}_{\bf x}}{\cal R}_i + {\bf u}$$
where ${\cal H}_{\bf x}$ is a subcode  of  ${\cal H}_{q,m}$  defined by a $m - 2$ dimensional hyperplane $supp({\bf x})$.
Therefore, minimum distance  graph of the  punctured code $({\cal R}_{i} + {\bf u})^i$ is a connected component of the minimum distance  graph of the punctured Hamming code ${\cal H}_{q,m}^i$.
\end{IEEEproof}

Following \cite{rom5},  we will call  a subspace ${\cal R}_i$ of   Hamming code ${\cal H}_{q,m}$   a principal $i$-component.

We now  present a generalization of the Lindstr\"{o}m \cite{lin} and Sch\"{o}nheim  \cite{sch} construction of $q$-ary 1-perfect  codes.


\begin{construction}
Let  ${\cal C}_1$ be a $q$-ary  1-perfect code of length $n = (q^m - 1) / (q - 1), \ m \geq 2$, let  ${\cal R}_i$ be a
principal $i$-component of the $q$-ary Hamming code ${\cal H}_{q,m + 1}$, $i \leq (q - 1)n + 1$, and let $\sigma_{\bf c}$ be a permutation of the elements in $\mathbb{F}_{q}$, ${\bf c} \in {\cal C}_1 $. Then we construct the code ${\cal C}$. Let
\begin{equation}
{\cal C} = \bigcup_{{\bf c} \in {\cal C}_1} \sigma_{\bf c}({\cal R}_i + ({\bf 0}|{\bf c}))
\end{equation}
where the zero vector ${\bf 0} \in \mathbb {F}_{q}^{(q - 1)n + 1}$.
\end{construction}


\begin{theorem}
The code ${\cal C}$ is a $q$-ary 1-perfect code of length $qn + 1$.
\end{theorem}

\begin{IEEEproof}

Consider a vector ${\bf x} \in  \mathbb{F}_{q}^{n}$ such that its $supp({\bf x})$ is $m - 2$ dimensional hyperplane. Without loss of generality we can assume that
$$supp({\bf x}) = \{(q - 1)n + 2, (q - 1)n + 3, \ldots,  q n + 1 \}. $$
Since $i \leq (q - 1)n + 1$, it follows that $i \notin  supp({\bf x})$
Hence taking into account that ${\cal R}_i$ is the principal $i$-component   of the $q$-ary Hamming code and $i \notin  supp({\bf x})$, we have that minimum distance  graph of the code $({\cal R}_i + ({\bf 0}|{\bf c}))^i$ is connected.  Lemma 1 implies  that
$$ ({\cal R}_i + ({\bf 0}|{\bf c}_1)) \cap  ({\cal R}_i + ({\bf 0}|{\bf c}_2)) = \varnothing$$
for all ${\bf c}_1, {\bf c}_2 \in {\cal C}_1$,  ${\bf c}_1 \ne {\bf c}_2$.
Further, we consider punctured codes $({\cal R}_{i} + ({\bf 0}|{\bf c}_1))^i $ and  $({\cal R}_{i} + ({\bf 0}|{\bf c}_2))^i$.
Since $i \notin  supp({\bf x})$, from Lemma 1 it follows that    for any ${\bf u}\in ({\cal R}_{i} + ({\bf 0}|{\bf c}_1))^i $ and for any ${\bf v}\in ({\cal R}_{i} + ({\bf 0}|{\bf c}_2))^i$  there holds the inequality $d({\bf u},{\bf v}) \geq 3$.
Hence  the code ${\cal R}_i + ({\bf 0}|{\bf c})$ is $i$-component of the code ${\cal C}$ for all ${\bf c} \in {\cal C}_1$.
The set of all $i$-components ${\cal R}_i + ({\bf 0}|{\bf c})$ form  a partition of the code ${\cal C}$  and  formula (2) takes the form (4).
The dimension of  ${\cal R}_{i}$ is $ q^{m} - 1$. Thus we have
$$|{\cal C}| = |{\cal R}_i|\cdot|{\cal C}_1| = q^{{q^{m}} - 1}\cdot q^{n - m} = q^{qn - m}.$$
Hence the number of codewords in the code ${\cal C}$ is correct.
Therefore  by Theorem 1, we obtain that the code ${\cal C}$ is a $q$-ary 1-perfect code of length $qn + 1$.
\end{IEEEproof}

Denote by $N(q,n)$  the number of  $q$-ary  1-perfect codes of length $n = (q^m - 1) / (q - 1)$. Then from the  Lindstr\"{o}m \cite{lin} and Sch\"{o}nheim  \cite{sch}  construction we have
$$ N(q,n) \geq {(q)}^{q^{\frac{n - 1}{q} - m - 1}}.$$
From (4) and from definition of principal $i$-component of  $q$-ary Hamming code  it follows that by permutation of elements in $\mathbb{F}_{q}$ we obtain different 1-perfect codes.
Hence we get   that
$$ N(q,n) \geq {(q!)}^{q^{\frac{n - 1}{q} - m - 1}}.$$

We now cite two theorems from \cite{rom6} which will be needed  in the next section.

\begin{theorem}[Romanov \cite{rom6}, Theorem 1]
Let a vector $ {\bf u} = (u_1, u_2, \dots, u_n) \in {\cal R}_i$ and a component $u_x$ of the vector $\bf u$ be nonzero, $x \neq i$. Then,  the line $l_{ix}$ has a point $y$ distinct from the points $i, x$ and such that component $u_y$ of the vector $\bf u$ is nonzero.
\end{theorem}


The next theorem follows directly from Theorem 4.


\begin{theorem}[Romanov \cite{rom6}, Theorem 2]
 Let $i \neq j$, a vector ${\bf u} = (u_1, u_2, \dots, u_n) \in ({\cal R}_i + {\cal R}_j)$, a component $u_x$ of the vector $\bf u$ be nonzero and the point $x$ does not lie on $l_{ij}$. Then,  the plane $P_{ijx}$ has a point  $y$ distinct from the points $i, j, x$ and such that component $u_y$ of the vector $\bf u$ is nonzero.
\end{theorem}


\section {Full-rank perfect codes}

Etzion and Vardy \cite{etz} proposed a switching construction of the full-rank binary 1-perfect codes.  They also  proposed an original method to construct an admissible family of $i$-components of the binary Hamming code and their construction of the full-rank binary 1-perfect codes is based on this method. In \cite{rom6}, the method of Etzion and Vardy   has been  generalized to $q$-ary codes. In this section, we present a generalization of the Etzion and Vardy construction of the full-rank binary 1-perfect codes  to 1-perfect codes over  finite  fields  of characteristic $2$.  This generalization is based   on results in \cite{rom6}. Finally in this section, we  present a modification of the Etzion and Vardy construction of the full-rank  1-perfect codes  for  finite  fields  of arbitrary characteristic.

So first we  give  a generalization of the  switching construction (Etzion and Vardy \cite{etz}) for  $q$-ary Hamming codes.

A family $\{ {\cal R}_{i_1} + {\bf u}_1,  {\cal R}_{i_2} + {\bf u}_2, \ldots, {\cal R}_{i_t} +  {\bf u}_t\}$ of  $i$-components (where $i \in \{ i_1, i_2, \ldots, i_t\}$) of a $q$-ary  Hamming    code $ {\cal H}_{q,m}$  is called {\it admissible} if for any $r, s \in \{ 1, 2, \ldots, t \}$, $r \ne s$, we have $({\cal R}_{i_{r}} + {\bf u}_r) \cap ({\cal R}_{i_{s}} + {\bf u}_s )= {\varnothing}$.

\begin{construction}
Let  $\{ {\cal R}_{i_1} + {\bf u}_1,  {\cal R}_{i_2} + {\bf u}_2, \ldots, {\cal R}_{i_t} +  {\bf u}_t\}$  be an admissible family of $i$-components of a $q$-ary  Hamming    code $ {\cal H}_{q,m}$ of length $n = (q^m - 1) / (q - 1)$, $i \in \{ i_1, i_2, \ldots, i_t\}$ and  let  $\sigma_{1},  \sigma_{2}, \ldots, \sigma_{t}$ be $t$ permutations of the elements in $\mathbb{F}_{q}$. Then we construct the code ${\cal C}$. Let
\begin{equation}
{\cal C} = \left({\cal H}_{q,m} \setminus  \bigcup_{s = 1}^t {\cal R}_{i_s} + {\bf u}_s \right) \bigcup \left( \bigcup_{s = 1}^t \sigma_{s}({\cal R}_{i_s}+ {\bf u}_s) \right).
\end{equation}
\end{construction}


\begin{theorem}
The code ${\cal C}$ is a $q$-ary 1-perfect code of length $n = (q^m - 1) / (q - 1)$.
\end{theorem}

\begin{IEEEproof}

We need to prove that the number of codewords in the code ${\cal C}$ is correct and minimum distance $d({\cal C})$ of the code ${\cal C}$ is equal to $3$. Obviously that $|{\cal R}_{i_s}|= |\sigma_{s}({\cal R}_{i_s})|$. Since the family $\{ {\cal R}_{i_1} + {\bf u}_1,  {\cal R}_{i_2} + {\bf u}_2, \ldots, {\cal R}_{i_t} +  {\bf u}_t\}$ of $i$-components is admissible,  it follows from (5) that
$$|{\cal C}| = |{\cal C}_1| - \sum_{s = 1}^t|{\cal R}_{i_s}| +  \sum_{s = 1}^t|\sigma_{s}({\cal R}_{i_s})| = |{\cal C}_1|
= q^{n - m}. $$
Further, we show that $d({\cal C}) = 3$.
The permutation $\sigma_{s}$ defines  an isometric transformation on the space $\mathbb{F}_{q}^n$. Thus we have
$d(\sigma_{s}({\cal R}_{i_s})) = 3$, $s \in \{1, 2, \ldots, t\}$.

Consider an $i$-component  ${\cal R}_{i} + {\bf u}$ and  $j$-component ${\cal R}_{j} + {\bf v}$ from $\{ {\cal R}_{i_1} + {\bf u}_1,  {\cal R}_{i_2} + {\bf u}_2, \ldots, {\cal R}_{i_t} +  {\bf u}_t\}$. We suppose that $i \ne j$. (If $i = j$ then see Theorem 1.) Without loss of generality  assume that $i \notin supp(\bf x)$ and  $j \notin supp(\bf x)$ where $supp(\bf x)$ is $m - 2$ dimensional hyperplane,
${\bf x} \in  \mathbb{F}_{q}^{n}$. From the conditions of the theorem it follows that $({\cal R}_{i} + {\bf u}) \cap ({\cal R}_{j}+ {\bf v}) = \varnothing$.  Further, we consider punctured codes $({\cal R}_{i} + {\bf u})^i$ and $({\cal R}_{j}+ {\bf v})^j$. Since
$i \notin supp(\bf x)$ and  $j \notin supp(\bf x)$, we obtain from Lemma 1 that for any ${\bf c}\in ({\cal R}_{i}+ {\bf u})^{i}$ and for any ${\bf c}'\in ({\cal R}_{j}+ {\bf v})^{j}$ there holds the inequality $d({\bf c},{\bf c}') \geq 3$. Therefore $d({\cal C}) = 3$.
\end{IEEEproof}

Next, we construct the vectors ${\bf c}_1, {\bf c}_1, \ldots, {\bf c}_m $ and show that these vectors are the codewords of the Hamming code ${\cal H}_{q,m}$ of length $n = q^{m}-1/q-1$.

In the parity-check matrix $H = \left[\bf h_1, \bf h_2, \dots, \bf h_n\right]$ of the Hamming code ${\cal H}_{q,m}$, we choose $m$ linearly independent columns. Without loss of generality  assume that we have chosen the columns  $\bf h_1, \bf h_2, \dots, \bf h_m$.

For each $ {\bf z} \in \mathbb {F}_{q}^{m}\setminus\{{\bf 0}\}$  there exists  a unique scalar $\alpha \in  \mathbb{F}_{q} \setminus \{0\}$ and a unique vector  column ${\bf h}_i \in H$ such that ${\bf z} = \alpha{\bf h}_i^T$.
Define a mapping $\xi$ from the nonzero vectors of $\mathbb {F}_{q}^{m}$    onto the vectors of weight $1$
in $\mathbb {F}_{q}^{n}$ as follows:
$$\forall {\bf z} \in \mathbb {F}_{q}^{m}\setminus\{{\bf 0}\},  \ \xi({\bf z}) = (x_1, x_2, \ldots, x_n )\in \mathbb {F}_{q}^{n} $$
\[\mbox{where  } x_i =  \left\{ \begin{array}{rl}
\alpha, & \mbox{if  } {\bf z} = \alpha {\bf h}_i^T, \\
0, & \mbox{if  } {\bf z} \neq \alpha {\bf h}_i^T.
\end{array} \right. \]
From now on, we will use the notation $\xi({\bf h}_i)$ instead of the notation $\xi({\bf h}_i^T)$.
Following \cite{etz}, we  define
\setlength{\arraycolsep}{0.0em}
\begin{eqnarray}
\nonumber {\bf c}_1&{} = {} &\xi({\bf h}_1) +  \xi({\bf h}_1 + {\bf h}_2 + {\bf h}_3) + \xi({\bf h}_1 + {\bf h}_2 + {\bf h}_4) \\
\nonumber & &  + \, \xi({\bf h}_1 + {\bf h}_3 + {\bf h}_4),\\
\nonumber {\bf c}_2& = &\xi({\bf h}_1) + \xi({\bf h}_2) + \xi({\bf h}_1 + {\bf h}_3 + {\bf h}_4)  \\
\nonumber & &  + \, \xi({\bf h}_2 + {\bf h}_3 + {\bf h}_4),\\
\nonumber {\bf c}_4& = &\xi({\bf h}_1) + \xi({\bf h}_2) +  \xi({\bf h}_3) + \xi({\bf h}_4)  \\
\nonumber & & + \, \xi({\bf h}_1 + {\bf h}_2 + {\bf h}_3) + \xi({\bf h}_1 + {\bf h}_2 + {\bf h}_4)  \\
\nonumber &   & + \,  \xi({\bf h}_1 + {\bf h}_3 + {\bf h}_4) + \xi({\bf h}_2 + {\bf h}_3 + {\bf h}_4).
\end{eqnarray}
\setlength{\arraycolsep}{5pt}
Further let  $j \in \{1, 2, \ldots, m \}\setminus \{1, 2, 4 \}$. If $j$ is odd, define
\setlength{\arraycolsep}{0.0em}
\begin{eqnarray}
\nonumber {\bf c}_j&{} = {}& \sum_{i = 1}^j \xi({\bf h}_i) +  \xi({\bf h}_1 + {\bf h}_2 +  \cdots + {\bf h}_j).
\end{eqnarray}
\setlength{\arraycolsep}{5pt}
Otherwise set
\setlength{\arraycolsep}{0.0em}
\begin{eqnarray}
\nonumber {\bf c}_j =  \sum_{i = 1}^j{} &\xi({\bf h}_i)&  {}+{}  \xi({\bf h}_1 + {\bf h}_2 +  \cdots + {\bf h}_{j/2})  \\
\nonumber &   &   \quad  \;  + \, \xi({\bf h}_{j/2 + 1} + {\bf h}_{j/2 + 2} +  \cdots + {\bf h}_j).
\end{eqnarray}
\setlength{\arraycolsep}{5pt}

\begin{lemma} If a finite field $\mathbb {F}_{q}$ has  characteristic $2$ and $m \geq 4$. Then, ${\bf c}_i \in {\cal H}_{q,m}$, $i = 1, 2, \ldots, m$.
\end{lemma}

\begin{IEEEproof}
Since the finite field $\mathbb {F}_{q}$ has  characteristic $2$ it follows by construction of ${\bf c}_i$ that ${\bf c}_i \in {\cal H}_{q,m}$, $i = 1, 2, \ldots, m$.
\end{IEEEproof}

Theorem 2 implies that ${\cal R}_{i} + {\bf u}$  is an  $i$-component of the $q$-ary  Hamming code ${\cal H}_{q,m}$ if ${\bf u} \in {\cal H}_{q,m}$. Therefore
by Lemma 1  it is clear that the family
$$ {\cal F} = \{ {\cal R}_1 + {\bf c}_1,  {\cal R}_2 + {\bf c}_2, \ldots, {\cal R}_m + {\bf c}_m \}$$
is a  family of $i$-components  of the Hamming code ${\cal H}_{q,m}$.

Further we  study the properties of the family
${\cal F}$
of $i$-components  of the Hamming code ${\cal H}_{q,m}$.


\begin{lemma}
 If finite field $\mathbb {F}_{q}$ has  characteristic $2$ and $m \geq 4$. Then, the family
${\cal F}$
of $i$-components   is an admissible family of $i$-components of the Hamming code ${\cal H}_{q,m}$ over the   finite  field $\mathbb {F}_{q}$.
\end{lemma}

\begin{IEEEproof}
Let $r,s \in \{1, 2, \ldots, m \}$, $r \ne s$. Then, we show that
\begin{equation}
({\cal R}_{{r}} + {\bf c}_{r}) \cap ({\cal R}_{{s}} + {\bf c}_{s}) = {\varnothing}.
\end{equation}
In order to satisfy  the equality 6, it suffices to show that
$${\bf c}_r - {\bf c}_s \notin {\cal R}_{r} + {\cal R}_{s}.$$
By Theorem 5, it suffices to show that  the support of  vector $ {\bf c}_r - {\bf c}_s $ contains a point $x$, not lying on the line $l_{r s}$ and such that no other point (distinct from the points  $ r,  s, x $) in the $supp({\bf c}_r - {\bf c}_s )$ does not belong to the plane $P_{rsx}$.
We consider several cases.

{\it Case  1:}
Let $r = 1$, $s = 2$.
Then
\setlength{\arraycolsep}{0.0em}
\begin{eqnarray}
\nonumber  & supp({\bf c}_1  - {\bf c}_2 ) = \{2\}\cup supp(\xi({\bf h}_1 + {\bf h}_2 + {\bf h}_3)) \phantom{aaaaaaaa}& \\
\nonumber &  \phantom{aaaaaaaaaaaaaaaaaaaaaaaaaa}  \cup supp(\xi({\bf h}_1 + {\bf h}_2 + {\bf h}_4)).&
\end{eqnarray}
\setlength{\arraycolsep}{5pt}
Suppose that $x = supp(\xi({\bf h}_1 + {\bf h}_2 + {\bf h}_3))$. Since the columns  $\bf h_1, \bf h_2, \dots, \bf h_m$  are linearly independent it is obvious that $({\bf h}_1 + {\bf h}_2 + {\bf h}_4)\notin P_{12x}$.

The cases when $r = 1$ or $r = 2$ and $s =3$ or $s = 4$ are proved similarly.

{\it Case  2:}
Let  $r,s \in \{1, 2, \ldots, m \} \setminus \{1, 2, 4 \}$, $r \ne s$ and $r, s$ is odd.
Without loss of generality we can assume that $r < s$. Then we have
\setlength{\arraycolsep}{0.0em}
\begin{eqnarray}
\nonumber & supp({\bf c}_r - {\bf c}_s ) = \{r + 1, r + 2, \ldots, s\} \phantom{aaaaaaaaaaaaaaaa} &\\
\nonumber&  \phantom{aaaaaaaaaa}  \bigcup supp\Bigl( \xi \Bigl(\sum_{i = 1}^r {\bf h}_i \Bigr) \Bigr) \bigcup supp\Bigl( \xi \Bigl(\sum_{i = 1}^s {\bf h}_i \Bigr) \Bigr).&
\end{eqnarray}
\setlength{\arraycolsep}{5pt}
Suppose that $x = r +1$. Then for any
$$ y \in supp({\bf c}_r - {\bf c}_s )\setminus \{r,  s, r + 1\}$$
 it is clear that $y \notin P_{rsr + 1}$.

{\it Case  3:}
Let  $r,s \in \{1, 2, \ldots, m \} \setminus \{1, 2, 4 \}$, $r \ne s$ and $r, s$ is even.
Without loss of generality we assume that $r < s$. Then we have
\setlength{\arraycolsep}{0.0em}
\begin{eqnarray}
\nonumber & supp({\bf c}_r - {\bf c}_s ) = \{r + 1, r + 2, \ldots, s\} \phantom{aaaaaaaaaaaaaaaa} &\\
\nonumber&  \phantom{a}  \bigcup supp\Bigl( \xi \Bigl(\sum_{i = 1}^{r/2} {\bf h}_i \Bigr) \Bigr) \bigcup supp\Bigl( \xi \Bigl(\sum_{i = {r/2 + 1}}^{r} {\bf h}_i \Bigr) \Bigr)&\\
\nonumber&  \phantom{aaaaaaaa}  \bigcup supp\Bigl( \xi \Bigl(\sum_{i = 1}^{s/2} {\bf h}_i \Bigr) \Bigr) \bigcup supp\Bigl( \xi \Bigl(\sum_{i = {s/2 + 1}}^s {\bf h}_i \Bigr) \Bigr).&
\end{eqnarray}
\setlength{\arraycolsep}{5pt}
Suppose that   $ x = supp\left(\xi\left(\sum_{i = 1}^{r/2} {\bf h}_i\right)\right)$. Then for any
$$ y \in supp({\bf c}_r - {\bf c}_s )\setminus \{r,  s, x\}$$
it is clear that $y \notin P_{rsx}$.

{\it Case  4:}
Let  $r,s \in \{1, 2, \ldots, m \} \setminus \{1, 2, 3, 4 \}$, $r \ne s$,   $r$ is even and $s$ is odd.
Without loss of generality we assume that $r < s$. Then we have
\setlength{\arraycolsep}{0.0em}
\begin{eqnarray}
\nonumber & supp({\bf c}_r - {\bf c}_s ) = \{r + 1, r + 2, \ldots, s\} \phantom{aaaaaaaaaaaaaaaa} &\\
\nonumber&  \phantom{a}  \bigcup supp\Bigl( \xi \Bigl(\sum_{i = 1}^{r/2} {\bf h}_i \Bigr) \Bigr) \bigcup supp\Bigl( \xi \Bigl(\sum_{i = {r/2 + 1}}^{r} {\bf h}_i \Bigr) \Bigr)&\\
\nonumber&  \phantom{aaaaaaaaaaa}  \phantom{\bigcup supp\Bigl( \xi \Bigl(\sum_{i = 1}^{s/2} {\bf h}_i \Bigr) \Bigr)} \bigcup supp\Bigl( \xi \Bigl(\sum_{i = 1}^s {\bf h}_i \Bigr) \Bigr).&
\end{eqnarray}
\setlength{\arraycolsep}{5pt}
Suppose that   $ x = supp\left(\xi\left(\sum_{i = 1}^{r/2} {\bf h}_i\right)\right)$. Then for any
$$ y \in supp({\bf c}_r - {\bf c}_s )\setminus \{r,  s, x\}$$
it is clear that $y \notin P_{rsx}$.

{\it Case  5:}
Let $ r = 1$,  $s \in \{1, 2, \ldots, m \} \setminus \{1, 2,  3,  4 \}$  and $s$ is odd.
\setlength{\arraycolsep}{0.0em}
\begin{eqnarray}
\nonumber & supp({\bf c}_r - {\bf c}_s ) = \{2, 3, \ldots, s\}\bigcup supp(\xi({\bf h}_1 + {\bf h}_2 + {\bf h}_3)) \phantom{aa} &\\
\nonumber&  \phantom{aaa}  \bigcup supp(\xi({\bf h}_1 + {\bf h}_2 + {\bf h}_4))\bigcup supp(\xi({\bf h}_1 + {\bf h}_3 + {\bf h}_4))\phantom{}  &\\
\nonumber&  \phantom{aaaaaaaaaaaaaaaaaaaaaaaaaaaaa} \bigcup supp\Bigl( \xi \Bigl(\sum_{i = 1}^s {\bf h}_i \Bigr) \Bigr).&
\end{eqnarray}
\setlength{\arraycolsep}{5pt}
Suppose that   $ x = s - 1$. Then for any
$$ y \in supp({\bf c}_r - {\bf c}_s )\setminus \{r,  s, x\}$$
it is clear that $y \notin P_{rsx}$.
The cases when $r = 1, 2 , 4$  and $s$ is odd  are proved similarly.

{\it Case  6:}
Let $ r =1$, $s \in \{1, 2, \ldots, m \} \setminus \{1, 2,  4 \}$  and $s$ is even.
Let $ r = 1$,  $s \in \{1, 2, \ldots, m \} \setminus \{1, 2,  3,  4 \}$  and $s$ is odd.
\setlength{\arraycolsep}{0.0em}
\begin{eqnarray}
\nonumber & supp({\bf c}_r - {\bf c}_s ) = \{2, 3, \ldots, s\}\bigcup supp(\xi({\bf h}_1 + {\bf h}_2 + {\bf h}_3)) \phantom{aa} &\\
\nonumber&  \phantom{aaa}  \bigcup supp(\xi({\bf h}_1 + {\bf h}_2 + {\bf h}_4))\bigcup supp(\xi({\bf h}_1 + {\bf h}_3 + {\bf h}_4))\phantom{}  &\\
\nonumber&  \phantom{aaaaaaaa} \bigcup supp\Bigl( \xi \Bigl(\sum_{i = 1}^{s/2} {\bf h}_i \Bigr) \Bigr) \bigcup supp\Bigl( \xi \Bigl(\sum_{i = {s/2 + 1}}^s {\bf h}_i \Bigr) \Bigr).&
\end{eqnarray}
\setlength{\arraycolsep}{5pt}
Suppose that   $ x = s - 1$. Then for any
$$ y \in supp({\bf c}_r - {\bf c}_s )\setminus \{r,  s, x\}$$
it is clear that $y \notin P_{rsx}$.
The cases when $r = 1, 2 , 4$  and $s$ is even  are proved similarly.
\end{IEEEproof}

\begin{theorem}
Let  $\mathbb {F}_{q}$  be a finite   field  of characteristic $2$, let ${\cal F} = \{{\cal R}_1 + {\bf c}_1,  {\cal R}_2 + {\bf c}_2, \ldots, {\cal R}_m + {\bf c}_m \}$ be the
family of $i$-components  defined above, $m \geq 4$, and  let  $\sigma_{1},  \sigma_{2}, \ldots, \sigma_{m}$ be $m$ permutations of the elements in $\mathbb{F}_{q}$, $ \sigma_i(1)\ne1$ for all  $1 \leq i \leq m$. Then the code
$$ {\cal C} = \left (  {\cal H}_{q,m} \setminus \bigcup_{i=1}^m  {\cal R}_{i} + {\bf c}_i  \right )  \bigcup \left (\bigcup_{i=1}^m  \sigma_i({\cal R}_{i} + {\bf  c}_i) \right) $$
is a full-rank 1-perfect code over the finite field  of characteristic $2$.
\end{theorem}


\begin{IEEEproof}
Lemma 3 implies that the family ${\cal F}$   is an admissible family of $i$-components of the Hamming code ${\cal H}_{q,m}$. Therefore by Theorem 6 we obtain that the code ${\cal C}$ is a $q$-ary 1-perfect code of length  $n = q^{m}-1/q-1$. From the definition of a Hamming code $ {\cal H}_{q,m}$ it follows that $rank( {\cal H}_{q,m})= n - m$.
The dimension of  ${\cal R}_{i}$ is $ q^{m-1} - 1$. Taking into account that
\setlength{\arraycolsep}{0.0em}
\begin{eqnarray}
\nonumber &\left| {\cal H}_{q,m} \setminus \bigcup_{i=1}^m  {\cal R}_{i} + {\bf c}_i \right|  \phantom{aaaaaaaaaaaaaaaaaaaaaaa} &\\
\nonumber &\phantom{aaaaaa}  =  q^{n - m} - mq^{q^{m-1}-1} >  \frac {1}{q}\cdot q^{n-m} = \frac {1}{q}\cdot \left| {\cal H}_{q,m} \right|.&
\end{eqnarray}
\setlength{\arraycolsep}{5pt}
We have $$rank\left({\cal H}_{q,m} \setminus \bigcup_{i=1}^m  {\cal R}_{i} + {\bf c}_i \right) = n-m. $$
By construction of  vectors ${\bf c}_1,  {\bf c}_2, \ldots, {\bf c}_m$, we have that the vectors
$${\bf c}_1(1,\sigma_1),  {\bf c}_2(2,\sigma_2), \ldots, {\bf c}_m(m,\sigma_m)$$ are
linearly independent. We remind that, by definition,
$$ {\bf c}(i, \sigma) = (c_1, c_2, \ldots, \sigma(c_i), \ldots, c_n). $$
Since $ \sigma_i(1)\ne1$ for all  $1 \leq i \leq m$, we have
$$ \{{\bf c}_1(1,\sigma_1),  {\bf c}_2(2,\sigma_2), \ldots, {\bf c}_m(m,\sigma_m)\} \cap {\cal H}_{q,m} = \varnothing .$$
But
$$ \{{\bf c}_1(1,\sigma_1),  {\bf c}_2(2,\sigma_2), \ldots, {\bf c}_m(m,\sigma_m)\} \subset {\cal C}.$$
Therefore, $rank({\cal C}) = n. $
\end{IEEEproof}


Finally, we  present a modification of the Etzion and Vardy construction of the full-rank  1-perfect codes  for  finite  fields  of arbitrary characteristic.
We define
\begin{eqnarray}
\nonumber {\bf c}_1& = &\xi({\bf h}_1) +  \xi({\bf h}_1 + {\bf h}_2 + {\bf h}_3) - \xi({\bf h}_1 + {\bf h}_2 - {\bf h}_4) - \\
\nonumber & & \xi({\bf h}_1 + {\bf h}_3 + {\bf h}_4),\\
\nonumber {\bf c}_2& = &\xi({\bf h}_1) + \xi({\bf h}_2) - \xi({\bf h}_1 - {\bf h}_3 - {\bf h}_4) - \\
\nonumber & & \xi({\bf h}_2 + {\bf h}_3 + {\bf h}_4),\\
\nonumber {\bf c}_4& = &\xi({\bf h}_1) - \xi({\bf h}_2) -  \xi({\bf h}_3) + \xi({\bf h}_4) + \\
\nonumber & & \xi({\bf h}_1 + {\bf h}_2 + {\bf h}_3) - \xi({\bf h}_1 + {\bf h}_2 + {\bf h}_4) - \\
\nonumber &   &   \xi({\bf h}_1 + {\bf h}_3 + {\bf h}_4) + \xi({\bf h}_2 + {\bf h}_3 + {\bf h}_4).
\end{eqnarray}
Further let  $j \in \{1, 2, \ldots, m \}\setminus \{1, 2, 4 \}$. If $j$ is odd, define
\begin{eqnarray}
\nonumber {\bf c}_j& = & \sum_{i = 1}^j \xi({\bf h}_i) -  \xi({\bf h}_1 + {\bf h}_2 +  \cdots + {\bf h}_j). \\
\nonumber \mbox {Otherwise set} \\
\nonumber {\bf c}_j& = & \sum_{i = 1}^j \xi({\bf h}_i)  -  \xi({\bf h}_1 + {\bf h}_2 +  \cdots + {\bf h}_{j/2}) - \\
\nonumber &   &   \qquad  \quad \xi({\bf h}_{j/2 + 1} + {\bf h}_{j/2 + 2} +  \cdots + {\bf h}_j).
\end{eqnarray}

Since the columns  $\bf h_1, \bf h_2, \dots, \bf h_m$ are linearly independent it follows  that ${\bf c}_i \in {\cal H}_{q,m}$, $i = 1, 2, \ldots, m$.  Further we define a  family
${\cal F} = \{ {\cal R}_{1} + {\bf c}_1,  {\cal R}_{2} + {\bf c}_2, \ldots, {\cal R}_{m} +  {\bf c}_m\}  $ of $i$-components  of the Hamming code ${\cal H}_{q,m}$.

\begin{theorem}
Let  $\mathbb {F}_{q}$  be a finite   field, let ${\cal F} = \{{\cal R}_1 + {\bf c}_1,  {\cal R}_2 + {\bf c}_2, \ldots, {\cal R}_m + {\bf c}_m \}$ be the
family of $i$-components  defined above, $m \geq 4$, and  let  $\sigma_{1},  \sigma_{2}, \ldots, \sigma_{m}$ be $m$ permutations of the elements in $\mathbb{F}_{q}$, $ \sigma_i(1)\ne1$ for all  $1 \leq i \leq m$. Then the code
$$ {\cal C} = \left (  {\cal H}_{q,m} \setminus \bigcup_{i=1}^m  {\cal R}_{i} + {\bf c}_i  \right )  \bigcup \left (\bigcup_{i=1}^m  \sigma_i({\cal R}_{i} + {\bf  c}_i) \right) $$
is a full-rank 1-perfect code over the finite field.
\end{theorem}

This theorem  is proved exactly like Theorem 7.

\section{ Conclusion}

Etzion and Vardy  \cite{etz} discovered a peculiar method  to construct an admissible family of $i$-components of the binary Hamming code. In \cite{rom6}, the method of Etzion and Vardy   has been  generalized to $q$-ary codes.
Besides the method of Etzion and Vardy, there is also method  to construct an admissible family of $i$-components of the  Hamming code suggested  in  \cite{rom1, rom2, rom3, mal, rom4}.
In \cite{rom1}, sufficient conditions for disjointness of $i$-components of  a binary Hamming code are obtained.
By means of these conditions, the  nonsystematic binary 1-perfect codes of length n = 15   were constructed, see \cite{rom2}. Also in \cite{rom3},  the  binary 1-perfect codes with trivial kernel   were constructed by means of the sufficient conditions from  \cite{rom1}.  In \cite{mal}, a criterion for disjointness of the $i$-components of a binary Hamming code is obtained, and  regular  partitions of the binary Hamming codes into $i$-components with new parameters are constructed.  In \cite{rom4}, the results from \cite{mal}  have been  generalized to $q$-ary codes.

Etzion  \cite{etz1}  generalized some results of \cite{etz} to $q$-ary case. Etzion  \cite{etz1} defined the $i$-components of a $q$-ary Hamming  code in terms of the generator matrices and  constructed $q$-ary 1-perfect codes by switching $i$-components of the $q$-ary Hamming  code for fixed coordinate $i$.

%







\end{document}